# Effect of Weather Conditions on FSO link based in Islamabad

Nauman Hameed*, Tayyab Mehmood Jatoi**, Habib Ullah Manzoor

*Department of Telecommunication, University of Engineering & Technology, Taxila, Pakistan,
nauman112te@gmail.com, **NUST, School of Electrical Engineering & Computer Science, Islamabad, Pakistan.

*Abstract*—Free space optics (FSO) is a field of curiosity and importance for the scientists because of its numerous applications and advantages like low cost FSO systems, easy deployment, high data rate, secure FSO links and license free bands. Very high bandwidth FSO link can be effectively established between the skyscrapers of the Islamabad Pakistan for the purpose high capacity applications in these skyscrapers. FSO links are badly affected by the weather conditions especially rain and fog because of high attenuation factors. OPTI-System is used to study the effect of rain and fog on the performance of FSO links.

## I. INTRODUCTION

In today's Internet community, demand of services consuming high data rates is increasing day by day. Research work is being carried out in the field of communication technology to fulfill high data rate demand with reliable quality of service and lowest cost possible. Security is a top priority in between communication of two or more parties. By taking all the facts into mind, Free Space Optics (FSO) is one of the choice to fulfill these demands. FSO provides data rates in Gbps in wireless scenario with the most secure communication because light beam is prone to eavesdropping. FSO is having many advantages over other wireless technologies like free licensing, ease of installation, low capital equipment cost and very high data rates. FSO was first used for military purposes. FSO has found fascinating applications in access networks (last mile solution), airborne & inter-satellite communication, disaster recovery, inter-building connections (point to point or multipoint) and short term installation for certain events [1]. FSO having larger bandwidth can be a solution to growing capacity hungry applications. Despite of its several merits, FSO may not be a good choice for many locations because FSO link is susceptible to bad weather conditions. FSO utilizes air as an interface for establishing link between transmitter and receiver. Therefore, weather conditions must be examined before practical implementation of FSO link. Factors which affect the FSO link are absorption, scintillation & scattering [2, 3]. Certain weather conditions like haze, rain, fog & snow have a different effect on optical transmission. The main factor which affects the most is fog. For maximum availability of an FSO link, it is necessary to evaluate these weather conditions over a large period of time before setting up a link. In this paper work, we have tried to estimate the performance of an FSO link based on weather conditions of Islamabad city. Optisystem is utilized to perform simulative analysis.

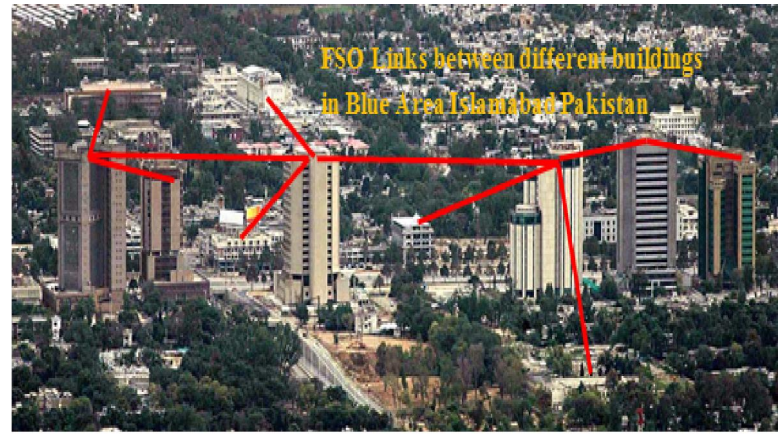

Fig. 1 Future of wireless communication, conceptual FSO links established in Blue Area Islamabad Pakistan

## II. SIMULATION SETUP

The system which is set up in Optisystem is shown in Fig. 2. First block shows a laser source having operating frequency of 1550nm because atmospheric attenuation produces less effect at this frequency. Second block shows a subsystem which consists of PRBS (Pseudo Random Bit Sequence) generator, NRZ pulse generator, low pass bessel filter and Machzender modulator as shown in Fig. 3. After that FSO channel is present which comprises of a link range of 1Km, attenuation factor, Tx and Rx aperture diameter and beam divergence. Fourth block is an APD (Avalanche Photo Detector) having ionization ratio 0.9 and 10nA dark current. Fifth block is a low pass bessel filter with cutoff frequency= 0.75*Bit rate and order equals to 4. At the end, BER analyzer is used to compute eye diagram, minimum BER and Q-factor of the designated system.

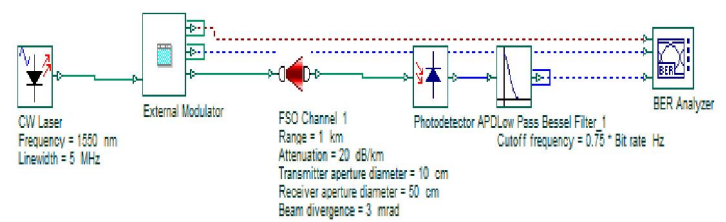

Fig. 2 Reference simulation model of FSO Link

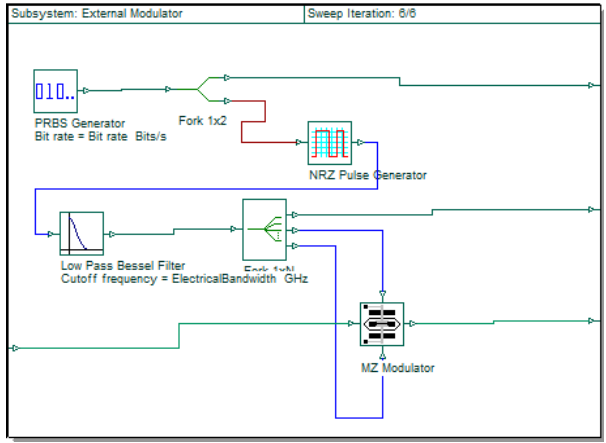

Fig. 3 Subsystem containing Bit sequence generator, NRZ pulse driver and Mach-zender modulator.

*1) Reference system characteristics*

| Design Parameters | Values |
|---|---|
| Data Rate | 10Gbps |
| Input Power | 5dBm |
| Link Range | 1Km |
| Operating Frequency | 1550nm |
| Modulator | Machzender Modulator |
| Sequence Length | 128 bits |
| Samples/bit | 64 |
| Number of Samples | 8192 |
| Beam Divergence | 2mrad |
| Optical Detector | Avalanche Photodiode |
| Filter Type | Low Pass Bessel filter |
| Filter Order | 4 |

### III. QUALITY OF FSO LINK

There are certain factors which decide how the FSO link works in specific environment. Detailed discussion is given as:

*A. Link Margin*

Link margin can be calculated by observing received signal power at the receiver side [4]. It is an important factor to be observed that can affect the quality of an FSO link. Mathematical expression for link margin (LM) is given as:

$$LM = 10 \log P_R / s \quad (1)$$

In the above equation, $P_R$ is a received signal power and **s** is receiver sensitivity. At receiver side, for signal to be detected its power should be greater than receiver sensitivity. Receiver sensitivity is a constant value in dBm given by manufacturer and ranging from -20 to -40dBm. So, received signal power must be evaluated for quality check of an FSO link.

*B. Geometric Attenuation*

Divergence of optical beam is caused due to geometric attenuation while propagating through air interface. It can be calculated for performance evaluation of an FSO link. The expression used for geometrical attenuation is given in [4] as:

$$A_{geo} = \left[\frac{d_{RX}}{d_{TX} + \theta l}\right]^2 \quad (2)$$

Where $d_{RX}$ and $d_{TX}$ are diameters of apertures for receiver and transmitter respectively. While '$\theta$' represents the divergence angle and 'l' denotes link length.

*C. Atmospheric Attenuation*

Attenuation which occurs in attenuation channel due to presence of aerosols is termed as atmospheric attenuation. As a result of atmospheric attenuation, light beam is partially distorted resulting in scattering, absorption and diffraction. From [5], expression is noted as:

$$\propto = e^{-\sigma l} \quad (3)$$

Whereas 'l' is distance between transmitter and receiver, '$\sigma$' is attenuation coefficient per unit length. Further value of '$\sigma$' can be calculated using Kim and Kruse relations.

Atmospheric attenuation produced due to the phenomenon of scattering and absorption of light beam can be calculated using Beer's Law [6] which is:

$$I = I_0 e^{(-\gamma x)} \quad (4)$$

Whereas '$I$' and '$I_0$' are detected and initial intensities at certain location 'x'. While '$\gamma$' is attenuation coefficient.

### IV. WEATHER INFLUENCE ON FSO LINK

Free Space Optics links are operated in open atmosphere, so local weather conditions and microphysics of atmosphere highly affect the propagating light signal. Certain parameters play their role in degradation of signal quality. Visibility determines that how far an optical signal can travel in open air. Various elements present in air can limit the visibility. Dust particles, smoke, rain, haze, fog and snow attenuate the signal at different intensities. Fog is the major attenuation factor because the size of its particles is similar to the wavelength of light used as a signal carrier. The size of snow particles is a bit larger therefore less attenuation is induced. In [7] the impact of different weather conditions like rain, fog and snow was investigated.

*A. Rain Attenuation*

Rain is one of the factor for inducing attenuation in a FSO system. Rain has less impact than fog because wavelength of optical signal is very small as compared to rain drop [6]. The attenuation of optical signal due to rain is by scattering phenomenon. Specific optical attenuation increases linearly with increase in rain rate. Rainy season in Islamabad starts at the end of June and remains till the end of September with average monsoon rainfall of 790.8mm [8]. In the months of December, January and February, mostly moderate rain is observed. The rain rate of 25mm/hr induces attenuation factor of 6dB/Km for 1Km of link length. FSO systems having 25dB of link margin are able to penetrate rain unobstructed [6]. Fig. 4 shows the attenuation factor vs. rain rate computed at Milan, Italy reported in [9]. For rain rates of 25, 50, 100, 150mm/hr, optical attenuation values are 7.3, 14.6, 23.8 and 30.38dB/Km respectively. Statistics in [10] show that average rainfall in the month of July and August is 100mm/hr which yields the attenuation value of 30.38dB/Km. After performing simulation, Q-factor value of 25 and BER value of 1.33e-144 at attenuation factor of 30.38dB/Km is achieved. The eye diagram is shown

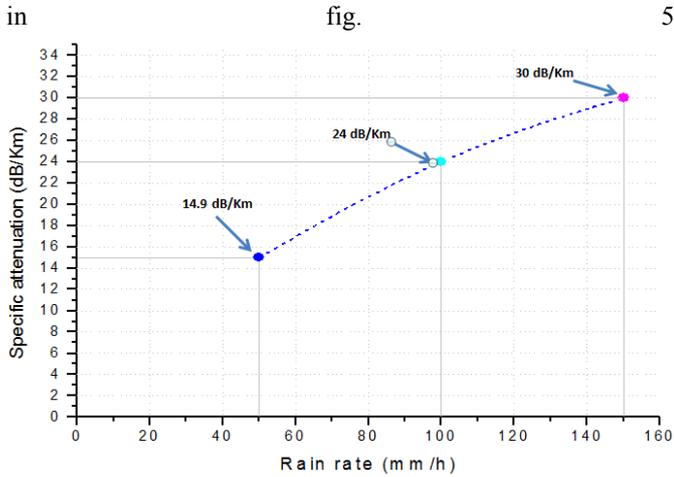
Fig. 4 Estimated optical attenuation at Milan, Italy.

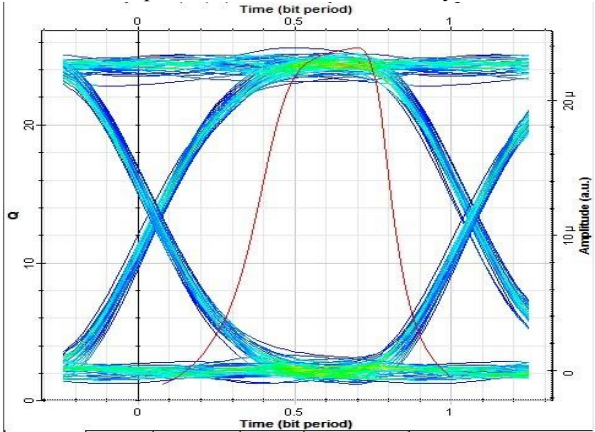
Fig. 5 Eye diagram for simulation setup of rain attenuation.

### B. Fog Effect

Fog is the most hostile factor to FSO link. Fog particles are having nearly the same wavelength as wavelength of light. Therefore, the attenuation caused by fog is extremely large with reference to other weather conditions. It is the most alarming weather condition for an FSO link. Fog events in Islamabad occur mostly in the months of November, December, January and February. In [11], four fog events are studied by using Kim, Kruse and Al Naboulsi model. Peak attenuation values for four events are 88.43, 77.88 and 110dB/Km for Kim, Kruse and Al Naboulsi model respectively. At such high value of attenuation, possible link length can be 500m for reliable communication. By taking link length of 500m with attenuation factor of 100dB/Km, simulation results show that Q-factor of 9.17 is achieved at BER value of 2.28e-20. Eye diagram for this scenario is shown in fig. 6.

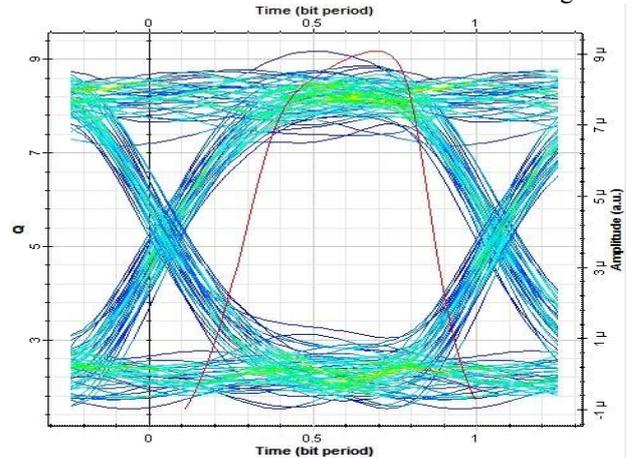
Fig. 6 Eye diagram for simulation setup of fog attenuation.

### V. POSSIBLE SCENARIOS OF FSO LINKS IN ISLAMABAD

Some possible scenarios for setting up FSO communication links are shown in table. 2 in the city of Islamabad. Google maps is used for capturing images and calculating distances between buildings. Aerial view of certain links in Islamabad is shown in fig. 7.

*2) Proposed FSO links*

| Link # | Transmitting building | Receiver building | Distance (Km) |
|---|---|---|---|
| 1 | Cisco Systems | KPMG Taseer Hadi & Co | 0.047 |
| 2 | Cisco Systems | Shaheed-e-Millat Secretariat | 0.132 |
| 3 | Shaheed-e-Millat Secretariat | OGDCL | 0.153 |
| 4 | OGDCL | United Bank Limited | 0.068 |
| 5 | OGDCL | Green Trust Tower | 0.103 |
| 6 | Green Trust Tower | HR Consultants | 0.522 |
| 7 | Green Trust Tower | NIC building | 1.452 |
| 8 | NIC building | Huawei technologies Pakistan | 0.191 |
| 9 | NIC building | State Life Tower | 0.064 |
| 10 | Huawei technologies Pakistan | Ufone Tower | 0.659 |
| 11 | Ufone Tower | Islamabad Stock Exchange | 0.050 |
| 12 | Ufone Tower | Centaurs | 0.851 |
| 13 | Centaurs | ZTBL | 1.728 |

### VI. CONCLUSION

Free Space Optics is a feasible solution for day-by-day increasing demands of bandwidth hungry applications. Exposure to bad weather conditions make its implementation nearly impossible for certain regions. For cities like Islamabad where weather conditions are reasonably bad for FSO links spanning certain kilometers, short links can be setup by calculating link margin and link budget with acceptable attenuation.

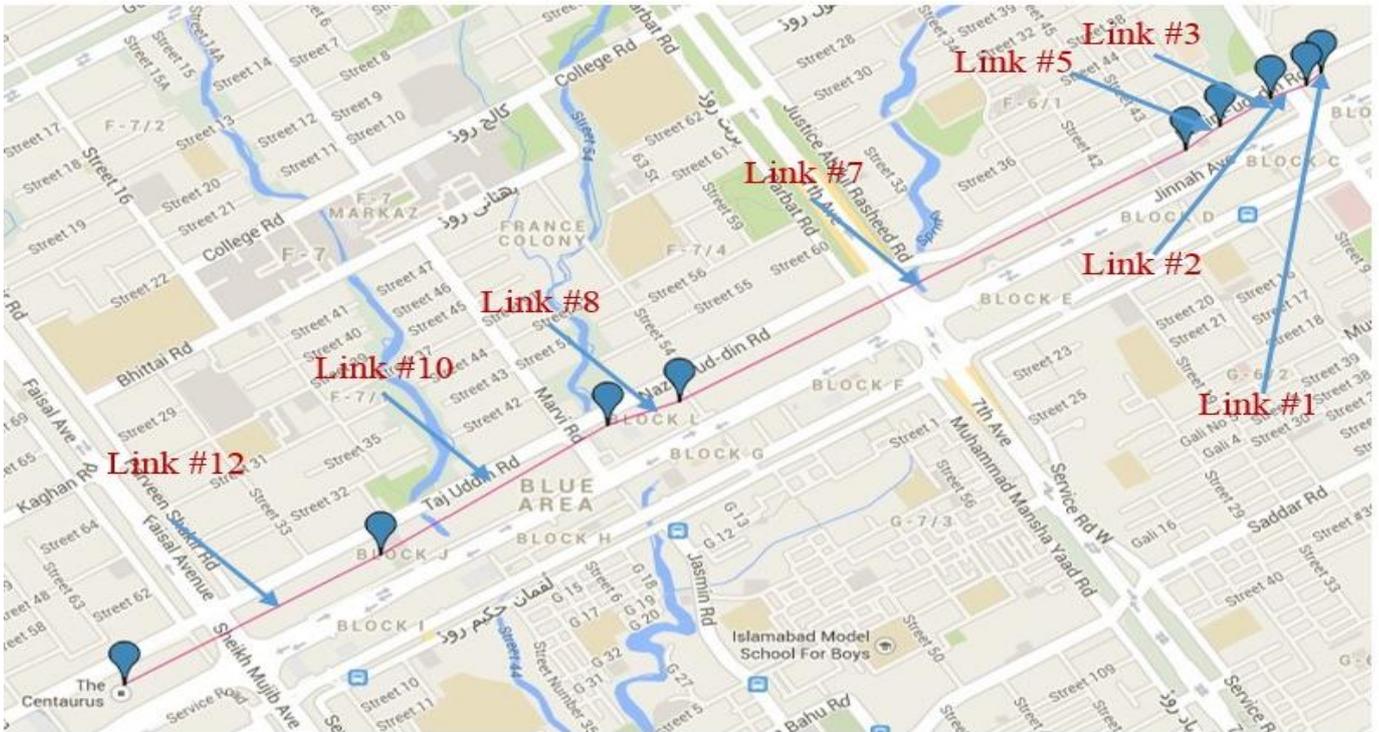

Fig. 7 Aerial view of the FSO links established between skyscrapers in Blue Area Islamabad